\documentclass{iaus}
\usepackage{graphicx,natbib}
\usepackage{wrapfig}

\title[Testing Relativistic Gravity from Space] 
{Tests of Relativistic Gravity from Space}

\author[Slava G. Turyshev]   
{Slava G. Turyshev}

\affiliation{Jet Propulsion Laboratory, California Institute of Technology,\\
4800 Oak Grove Drive, Pasadena, CA 91109-0899, USA \\ email: {\tt turyshev@jpl.nasa.gov}}

\pubyear{2009}
\volume{261}  
\pagerange{1--126}
\jname{Relativity in Fundamental Astronomy: \\
Dynamics, Reference Frames, and Data Analysis}
\editors{S. Klioner, P.\,K. Seidelmann and M. Soffel., eds.}
\begin{document}

\maketitle

\begin{abstract}
Recent experiments have successfully tested Einstein's general theory of relativity to remarkable precision. We discuss recent progress in the tests of relativistic gravity in the solar system and present motivations for the new generation of high-accuracy gravitational experiments. We especially focus on the concepts aiming to probe parameterized-post-Newtonian parameter $\gamma$
and evaluate the discovery potential of the recently proposed experiments.
\end{abstract}

\firstsection 

\section{Status of Precision Tests of Gravity}

Ever since its original publication in 1915, Einstein's  general theory of relativity  continues to be an active area of both theoretical and experimental research. Presently, the theory successfully accounts for all data gathered to date \cite[]{Turyshev-UFN:2009}.

To describe the accuracy achieved in the solar system gravitational experiments, 
it is useful to refer to the parameterized post-Newtonian (PPN) formalism \cite[]{Will-lrr-2006-3}. A particular metric theory of gravity in the PPN formalism with a specific coordinate gauge is fully characterized by means of ten PPN parameters.  The formalism uniquely prescribes the values of these parameters for the particular theory under study.\footnote{In a special case, when only two PPN parameters ($\gamma$, $\beta$) are considered, these parameters have a clear physical meaning. The parameter $\gamma$  represents the measure of the curvature of space created by a unit rest mass; parameter  $\beta$ represents a measure of the non-linearity of the law of superposition of the gravitational fields in a theory of gravity.} General relativity, when analyzed in the standard PPN gauge, gives $\gamma=\beta=1$; other theories may yield different values of these  parameters. Gravity experiments can be analyzed in terms of the PPN metric; an ensemble of experiments determine the unique value for these parameters and hence the metric field itself.

\begin{wrapfigure}{R}{0.53\textwidth}
  \vspace{-5pt}
  \begin{center}
    \includegraphics[width=0.53\textwidth]{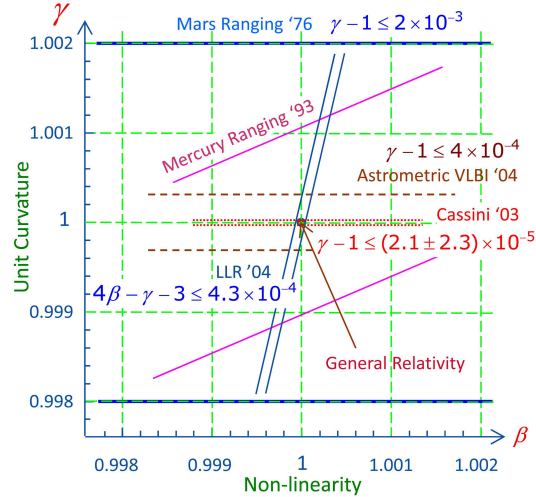}
  \end{center}
  \vspace{-0pt}
  \caption{Progress in the measurement accuracy of the Eddington's parameters $\gamma$ and $\beta$. So far, general theory of relativity survived every test \cite[]{Turyshev-UFN:2009}, yielding $\gamma-1=(2.1\pm2.3)\times10^{-5}$ \cite[]{Bertotti-Iess-Tortora-2003} and $\beta-1=(1.2\pm1.1)\times10^{-4}$ \cite[]{Williams-etal-2004}.}
\label{fig:gamma-beta}
  \vspace{-10pt}
\end{wrapfigure}

Over the years, a number of solar system experiments contributed to the phenomenological success of general relativity. Analysis of data obtained from microwave ranging to the Viking lander on Mars verified the prediction of this theory that the round-trip times of light signals traveling between the Earth and Mars are increased by the direct effect of solar gravity; the corresponding value of the PPN parameter $\gamma$ was obtained at the level of $1.000 \pm 0.002$ \cite[]{viking_reasen} (see Fig.~\ref{fig:gamma-beta}). Analyses of very long baseline interferometry data have yielded result of $\gamma=0.99983\pm0.00045$ \cite[]{Shapiro_SS_etal_2004}. Lunar laser ranging constrained a combination of PPN parameters $4\beta-\gamma-3=(4.0\pm4.3)\times10^{-4}$ 
via precision measurements of the lunar orbit \cite[]{Williams-etal-2004}.  
Finally, microwave tracking of the Cassini spacecraft on its approach to Saturn measured the parameter $\gamma$ as $\gamma-1=(2.1\pm2.3)\times10^{-5}$, thereby reaching the current best accuracy provided by tests of gravity in the solar system \cite[]{Bertotti-Iess-Tortora-2003}. 

It is remarkable that even more than 90 years after general relativity was conceived, Einstein's theory has survived every test \cite[]{Will-lrr-2006-3}.  
Such longevity and success make general relativity the de-facto ``standard'' theory of gravitation for all practical purposes involving spacecraft navigation and astrometry, astrophysics, cosmology and fundamental physics \cite[]{Turyshev-UFN:2009}. 

At the same time there are many important reasons to question the validity of general relativity and to determine the level of accuracy at which it is violated. On the theoretical front, problems arise from several directions, most concerning the strong-gravitational field regime. These challenges include the appearance of spacetime singularities and the inability of a classical description to describe the physics of very strong gravitational fields. A way out of this difficulty may be through gravity quantization. However, despite the success of modern gauge-field theories in describing the electromagnetic, weak, and strong interactions, we do not yet understand how gravity should be described at the quantum level. This continued inability to merge gravity with quantum mechanics indicates that the pure tensor gravity of general relativity needs modification or augmentation. 

In addition, recent remarkable progress in observational cosmology has subjected the general theory of relativity to increased scrutiny by suggesting a non-Einsteinian scenario of the Universe's evolution. Researchers now believed that new physics is needed to resolve these issues.
Theoretical models of the kinds of new physics that can solve the problems described above typically involve new interactions, some of which could manifest themselves as violations of the EP, variation of fundamental constants, modification of the inverse-square law of gravity at short distances, Lorenz symmetry breaking, or large-scale gravitational phenomena.   Each of these manifestations offers an opportunity for space-based experimentation and, hopefully, a major discovery.

Given the immense challenge posed by the unexpected discovery of the accelerated expansion of the universe, it is important to explore every option to explain and probe the underlying physics.  Theoretical efforts in this area offer a rich spectrum of new ideas, some discussed below, that can be tested by experiment.

Motivated by the dark-energy and dark-matter problems, long-distance gravity modification is one of the proposals that have recently gained attention \cite[]{Deffayet-Dvali-Gabadadze-2002}.  Theories that modify gravity at cosmological distances exhibit a strong coupling phenomenon of extra graviton polarizations \cite[]{Dvali-2006}. This phenomenon plays an important role in this class of theories in allowing them to agree with solar system constraints.  In particular, the ``brane-induced gravity'' model \cite[]{Dvali-Gabadadze-Porrati-2000} provides an interesting way of modifying gravity at large distances to produce an accelerated expansion of the universe, without the need for a non-vanishing cosmological constant \cite[]{Deffayet-Dvali-Gabadadze-2002}. One of the peculiarities of this model is means of recovering the usual gravitational interaction at small (i.e. non-cosmological) distances, motivating precision tests of gravity on solar system scales \cite[]{Dvali-Gruzinov-Zaldarriaga-2003,Magueijo-Bekenstein-2007}.   

\section{Future Space-based Tests of Gravitational Theories}
\label{sec:mod-grav-tests}

\begin{wrapfigure}{R}{0.27\textwidth}
  \vspace{-10pt}
  \begin{center}
    \includegraphics[width=0.27\textwidth]{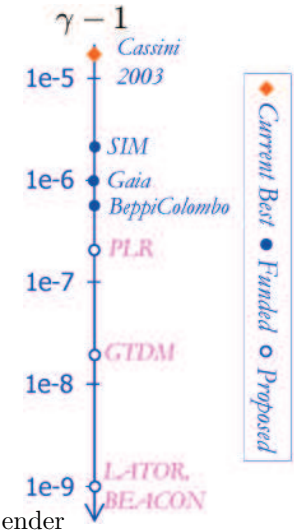}
  \end{center}
  \vspace{-0pt}
  \caption{Anticipated progress in the tests of the PPN parameter $\gamma$ \cite[]{Turyshev-UFN:2009}.}
  \vspace{-5pt}
\label{fig:future-gamma-tests}
\end{wrapfigure}

The Eddington parameter $\gamma$, whose value in general relativity is unity, is perhaps the most fundamental PPN parameter \cite[]{Will-lrr-2006-3}, in that $\frac{1}{2}(\gamma-1)$ is a measure, for example, of the fractional strength of the scalar-gravity interaction in scalar-tensor theories of gravity \cite[]{Damour_Nordtvedt_93b}. The current best accuracy of the Cassini\footnote{A similar experiment is planned for the ESA's {BepiColombo} mission to Mercury \cite[]{Iess-Asmar:2007} (see Fig.~\ref{fig:future-gamma-tests}).} $\gamma$ result of $\gamma -1 = (2.1\pm2.3)\times10^{-5}$ approaches the region where multiple tensor-scalar gravity models, consistent with the recent cosmological observations \cite[]{Spergel-etal-2006}, predict a lower bound for the present value of this parameter at the level of $\gamma-1 \sim 10^{-6}-10^{-7}$ \cite[]{Damour_Nordtvedt_93b,Damour_Piazza_Veneziano_02b}. Therefore, improving the measurement of this parameter would provide crucial information to separate modern scalar-tensor theories of gravity from general relativity, probe possible ways for gravity quantization, and test modern theories of cosmological evolution.

The current accuracy of modern optical astrometry, as represented by the Hipparcos Catalogue, is about 1 milliarcsecond, which gave a determination of $\gamma$ at the level of $0.997 \pm 0.003$ \cite[]{Froeschle-Mignard-Arenou-1997-gamma}. Future astrometric missions such as Gaia and the Space Interferometry Mission (SIM) will push the accuracy to the level of a few microarcseconds; the expected accuracy of determinations of the PPN parameter $\gamma$ will be $10^{-6}$ to $5\times10^{-7}$ \cite[]{Turyshev-SIM:2009}. 

Interplanetary laser ranging could lead to a significant improvement in the measurement accuracy of the PPN parameter $\gamma$. Thus, precision laser ranging between the Earth and a lander on Phobos (i.e., Phobos Laser Ranging -- PLR) during solar conjunctions may offer a suitable opportunity. If the lander were equipped with a laser transponder capable of reaching a precision of 1~mm, a measurement of $\gamma$ with accuracy of 2 parts in 10$^7$ would be possible. 

The Gravitational Time Delay Mission (GTDM) \cite[]{Ashby-Bender-2006} proposes to use laser ranging between two drag-free spacecraft (with spurious acceleration levels below $1.3 \times  10^{-13}~{\rm m/s}^2/\sqrt{\rm Hz}$ at  $0.4~\mu$Hz) to accurately measure the Shapiro time delay for laser beams passing near the Sun. One spacecraft will be kept at the L1 Lagrange point of the Earth-Sun system; the other one will be  placed on a 3:2 Earth-resonant orbit. A high-stability frequency standard ($\delta f/f\lesssim 1 \times  10^{-13}~{\rm Hz}^{-1/2}$ at $0.4~\mu$Hz) located on the L1 spacecraft will permit accurate measurement of the time delay. If requirements on the performance of the disturbance compensation system, the timing-transfer process, and the high-accuracy orbit determination are successfully addressed \cite[]{Ashby-Bender-2006}, then determination of the time delay of interplanetary signals to a 0.5 ps precision in terms of the instantaneous clock frequency could lead to an accuracy of 2 parts in $10^{8}$ in measuring parameter $\gamma$.

The Laser Astrometric Test of Relativity (LATOR) \cite[]{stanford_ijmpd} proposes to measure parameter $\gamma$ with an accuracy of 1 part in 10$^9$, which is a factor of 30,000 beyond the  best currently available, Cassini's 2003 result \cite[]{Bertotti-Iess-Tortora-2003}.  The key element of LATOR is a geometric redundancy provided by the long-baseline optical interferometry and interplanetary laser ranging. By using a combination of independent time-series of gravitational deflection of light in immediate proximity to the Sun, along with measurements of the Shapiro time delay on interplanetary scales (to a precision better than 0.01 picoradians and 3 mm, respectively), LATOR will significantly improve our knowledge of relativistic gravity and cosmology. LATOR's primary measurement, the precise observation of the non-Euclidean geometry of a light triangle that surrounds the Sun, pushes to unprecedented accuracy the search for cosmologically relevant scalar-tensor theories of gravity by looking for a remnant scalar field in today's solar system.  LATOR could lead to very robust advances in the tests of fundamental physics. It could discover a violation or extension of general relativity or reveal the presence of an additional long range interaction.

\begin{wrapfigure}{R}{0.520\textwidth}
  \begin{center}
    \includegraphics[width=0.520\textwidth]{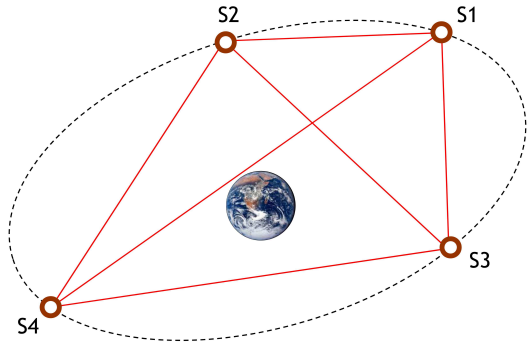}
  \end{center}
  \vspace{-0pt}
  \caption{Schematic of the BEACON formation.}
\label{fig:BEACON-formation}
  \vspace{-0pt}
\end{wrapfigure}

Similar to LATOR, the Beyond Einstein Advanced Coherent Optical Network (BEACON) \cite[]{Turyshev-etal-2008:BEACON} is an experiment designed to reach a sensitivity of 1 part in 10$^9$ in measuring the PPN parameter $\gamma$. 
The superior sensitivity of BEACON is enabled by redundant optical-truss architecture which eliminates the need for expensive drag-free systems.  The mission uses four identical commercially-available spacecraft that are placed on 80,000 km co-planar circular orbits around the Earth (Fig.~\ref{fig:BEACON-formation}). Each spacecraft is equipped with three sets of identical laser ranging transceivers to form a system of range measurements within the resulted flexible light-trapezoid formation. To enable its primary science objective, BEACON will precisely measure and monitor all six inter-spacecraft distances within the trapezoid using laser transceivers capable of achieving a nanometer resolution over distances of 160,000 km. 

In a planar geometry this system is redundant; by measuring only five of the six distances one can compute the sixth one.  The resulting geometric redundancy is the key element that enables BEACON's unique sensitivity in measuring a departure from Euclidean geometry.  In the Earth's vicinity, this departure is primarily due to the curvature of the relativistic space-time around the Earth. It amounts to $\sim$10 cm for laser beams just grazing the surface of the Earth and then falls off inversely proportional to the impact parameter. The BEACON's laser measurements form a trapezoid with diagonal elements such that one of the legs in the trapezoid skims close to the Earth, picking up an additional gravitational delay; the magnitude of this signal is modulated by moving the position of one of the spacecraft relative to the others (thus changing the impact parameter of the trapezoid legs). 

The BEACON architecture trades drag-free operation for redundancy in the optical truss. BEACON requires 4 Earth-orbiting satellites moving in the same orbital plane to participate in the metrology truss. This evolving light-trapezoid architecture is the fundamental requirement for BEACON; it enables geometric redundancy, thereby eliminating the need for drag-free spacecraft. Given range measurements of all 6 legs between the 4 fiducials, and assuming they are held in a planar configuration to within a few meters, it becomes possible to significantly improve the measurement of the PPN parameter $\gamma$. Simultaneous analysis of the resulting time-series of these distance measurements will allow BEACON to measure the curvature of the space-time around the Earth with an accuracy of better than 1 part in $10^9$ \cite[]{Turyshev-etal-2008:BEACON}.  

\section{Discussion and Outlook}
\label{sec:conclusions}

Today physics stands at the threshold of major discoveries.  Growing observational evidence points to the need for new physics, intensifying the efforts to resolve the challenges that fundamental physics faces today  \cite[]{Turyshev-UFN:2009}. We emphasize that modern-day optical technologies could  lead to important progress in the tests of relativistic gravity.

The experiments discussed above necessitate important modeling work needed to enable the anticipated high accuracy results \cite[]{Turyshev-SIM:2009}. The models would have to account for a number of relativistic-gravity effects on the light propagation in the solar system valid to the post-post-Newtonian order for a number of competing theories of gravity, including general relativity.  Special attention must be paid to the static gravitational fields produced by the mass monopoles of the Sun, Earth and Jupiter. The model should also be able to account for quadrupole deflection of light for the observations conducted in close proximity to the Sun and some planets. Lastly, gravitomagnetic effects due translational motion of the Sun and planets are important and should be accounted for. The  model would have to rely on a robust realization of the theory of relativistic reference frames, including relevant coordinate transformations and associated constants. 

The work outlined here will require a coordinated community effort to develop a reliable and effective set of relativistic modeling tools and data analysis algorithms. The work has already begun in the context of the development of relativistic reference frames and their application for precision spacecraft navigation, astrometry and gravitational experiments (the recent IAU Symposium 261 is a good evidence of such efforts), but more efforts are needed; this short paper was intended to motivate such a work in the near future.

The work described here was carried out at the Jet Propulsion Laboratory, California Institute of Technology, under a contract with the National Aeronautics and Space Administration.

\end{document}